# Preliminary Results on Vibration Damping Properties of Nanoscale-Reinforced Materials

*M.V. Kireitseu[1,2], G.R. Tomlinson[1], J. Lu[4], H. Altenbach[5], G. Rongong[1], L.V. Bochkareva[2], D. Hui[3]*

[1]Rolls-Royce Centre in Damping, Department of mechanical Engineering, University of Sheffield, UK
[2]UIIP National Academy of Sciences of Belarus, Minsk, Belarus
[3]Composite Nano/Materials Research Laboratory, University of New Orleans, USA
[4]Mechanical systems engineering department, University of Technology of Troyes, France
[5]Faculty of Engineering Sciences, Martin-Luther-Universitet Halle-Wittenberg, Germany

**ABSTRACT**

Our international team has recognized a role that nanotechnology-based solutions can have on dynamic/damping systems in areas of transportation (aerospace, automotive, railway and maritime). The novel concept of nanoparticle-based damping technology shows that a molecule-level mechanism can considerably enhance vibration damping and dynamic of machine components (such as fan blades or car dampers) via enhanced energy dissipation because of large surface-to-volume aspects in nanoparticle-reinforced composite material, large damping energy sources for friction and slip-stick motion at interfaces of matrix and nanoparticle. Thus carbon nanotube can act as a simple nanoscale damping spring in aerospace materials and is suggested for damping materials of the next generation. The materials offer the potential to further reduce the mass and dimension, increase performance, and reduce vibrations in wide-ranging applications. The above issues are being analyzed and reviewed in the paper by means of comprehensive and as complete as possible expert views. The current research works concentrate on an investigation related to nanoparticles/fibres/tubes-reinforced materials and coatings dynamic characterization and modeling of the fundamental phenomena that control relationships between structure and damping/mechanical properties of the materials.

## 1. INTRODUCTION

Vibrations and noise exist in almost every aspect of our life and are usually undesirable in engineering structures [1]. Vibrations are of concern in large structures such as civil (airbus A 380) or Rolls-Royce powered military aircrafts [2, 3], as well as small structures such as electronics [4]. Manufacturers have stated that for the next generation of transportation we need light-weight, cost-effective and reliable vibration-absorbing materials [5, 6]. The most important parameters that affect vibration damping in aerospace-related structures are 1) temperature (ranged from $-140^{\circ}$C in cryogenic applications and up to $600^{\circ}$C and higher), 2) rotational speed (up to 10.000 rpm in turbine engines), 3) frequency and amplitude of vibrations.

It is now accepted that nanotechnology may considerably enhance strength/damping behaviour and reduce noise of engineering structures through the utilisation of nanomaterials that dissipate a substantial fraction of the vibration energy that they receive [7]. Carbon nanotubes are particularly promising cost-decreasing reinforcement material [8, 13]. Boron nitride or silicon carbide nanotubes are another possible candidate for aerospace material reinforcement [2, 9]. The benefits that may be achieved by both carbon nanotubes in lower temperature damping applications and boron-nitride/silicon carbide nanotubes in high temperature aerospace applications may be very significant. The present paper will outline some preliminary efforts made by the institutions in this direction under EU programmes. The key goal of this communication is to provide a route map in the future research and outline a project concentrated on the next generation industry-oriented nanotechnology-based solutions for enhanced vibration damping/dynamics performance.

## 2. MANUFACTURING TECHNOLOGY, MATERIAL DESIGN AND APPLICATIONS

Selection of engineering material for the future nanoparticle-reinforced composite material is available from three types of solids that are metals, polymers and some ceramics. Selected matrix will mostly determine mechanism of energy dissipation [10, 11]; however, small CNT volume (1-5%) may greatly affect material's behaviour due to some extreme nanotube properties over traditional materials [5-12].

Nanotube-metal matrix composite materials are still rarely studied [6, 13]. Most used metallic alloys are hard or soft materials such as titanium/nickel and aluminium/bronze respectively. While titanium is stiff, light-weight and good for high temperature applications such as turbine fan blades, car panels are made of light-weight and cheaper aluminium and its alloys. Produced metal foams have excellent energy absorbing/damping properties over bulk material [14] and most likely are used for the next generation engineering. The CNT-reinforced metal materials are generally prepared by standard powder metallurgy melting [16] or ultrasonic liquid infiltration method [15], but good mixing and dispersion of the nanotubes should be achieved. In this respect it would be worth producing CNT-reinforced metallic foam composites and investigate their damping/dynamics performance.

Carbon nanotube-reinforced ceramic-matrix composite materials are a bit more frequently studied [17]; most successful efforts were made to obtain tougher ceramics (SiC, $Al_2O_3$, etc.) [37]. The composites can be processed by 1) mechanically mixed nanotubes with the matrix and then sintering [13, 37], melting [20] or spraying [18] of the particle mixture; 2) CVD deposited CNT-based thin films on SiC substrate (up to 50 μm) [19]; and 3) electro-chemical processing such as micro-arc oxidizing of metal substrate in an liquid electrolyte with added nanoparticles [21]. Some ceramic coatings are successfully used to enhance damping of titanium fan blades [18] and therefore, would be recommended as a candidate matrix material, but dispersion and orientation of nanotubes in the matrix is yet out of some control. Bulk properties other than mechanical are also worth being investigated.

Nanotube-polymer composites are now intensively studied [22-29], notably epoxy- and polymethylmethacrylate (PMMA)-matrix composites [25]; however, their damping behaviour is rather contradictory result than plausible information (see table 1). The ability of the polymer molecular chains to form large-diameter helices around individual nanotubes favours the formation of a strong bond with the matrix [36]. Selection of related manufacturing technologies is available from some well-known in the aircraft industry that are 1) manually or ultrasonically melt mixing and extrusion of nanotubes and polymer-layered silicate [12, 22], 2) CNT-reinforced resin by using so-called calandering technique [23, 24], 3) polymerization of carbon fibre by interfacial polymerization [26, 27].

Damping and integrity in nanocomposites have opposite requirements. Polymeric materials can provide enhanced damping over traditional metals and ceramics in normal environmental conditions, but their stiffness, performance and durability is decreased at high temperatures (above 200°C) [28] to be improved by nanoparticles. High damping and increased stiffness of a material is the best option, but it is difficult to achieve due to some uncertainties and poor fundamental knowledge on the added filler material. These require further optimization and tailoring for the next generation designs as a function of the number of parameters to be highlighted below.

**Table 1** Damping performance of CNT-reinforced polymeric materials

| Material | Young modulus at 25°C, GPa | Loss factor at 25°C |
|---|---|---|
| Epoxy resin Polyurethane at 850Hz Epoxy adhesive filler | 3.1-3.4 [12, 19] 0.3 [19] 3 [19, 34] and 0.00275 at 850Hz [19] | 0.01 [12]* 0.58 [19]** 0.4 [19] |
| 1-5 wt% CNT-matrix 50-60 wt% CNT-matrix | 3.2-3.6 [23, 24], 7.1-7.5 [35] | 0.08 [12] (* - 8 times increase) Not tested |
| MWNT-reinforced thin film (no matrix) at 850Hz | 284 [19] Bending stiffness + 30% [19, 34] | 0.3 [19] (** - 2 times decrease) +200% [19, 34] |

Size of reinforcement nanoparticles may vary from 1 to 100 nm in diameter and length. The number of CNT walls and their size affect on stress concentration in the composite [29] and thus short and even round particles are the strongest ones (diamonds etc.), but longer fibres are flexible and may be worth for damping while CNT may particularly act as a simple nanoscale spring and a crack trapping nanomaterial blocking the holes in the composites [30, 31]. Such a damping phenomenon could be multiplied by a factor of billions when CNTs are dispersed in a material.

Orientation and geometry (waviness) of CNT particles may affect mechanisms of energy dissipation/fracture mechanics and maximum stiffness is achievable at $90^0$ longitudinal CNT orientations [25, 29]. Notably those open-end CNTs do not collapse/failure/buckle due to higher stress concentrations while many authors [12, 22-31] have used closed-end CNT-reinforced composites. Thus isolated single-walled open-end nanotubes (SWNT) may be desirable for the future damping applications due to significant load-bearing ability in the case of CNT-matrix interactions. Defects of carbon particles likely limit the performance [23-27]. In a future CNT may serve as a storage container for sealant or multi-purpose particles of another material [13, 24]. CNT dispersion should be optimized for damping at 1-5% volume content because of carbon fibre conglomeration at higher volumes [32, 35], but 60 wt% CNT concentration in polymer was modelled [35]. The main disadvantage of CNT dis-persion is that it involves a large uncertainty to desired damping effect due to nanoparticles. Dynamic and damping properties of such the composite material are still at the beginning stage and CNT-phenomenon in vibration damping/dynamics should be investigated as a function of temperature, amplitude and frequency across the length scale.

Topics that could yield particular success for damping in the future include

1. Micro and nano-scale damping materials: polymer materials show the effect that extra energy dissipating motion of nanoparticles/molecules can have on the damping, strength and dynamics of material. Molecule-level mechanisms could give novel properties and nanoscale reinforcement might show novel damping phenomenon. A foam material based on nanoparticle dampers might give exceptional, temperature independent damping and negligible added weight.

2. Optimisation methods: Most damping systems require optimisation. Current material models [5-7, 12-19, 25-36] are too cumbersome to allow damped systems to be optimised effectively. Critical issues related to nanoparticle-reinforced materials include need to develop an affordable, reproducible computational methods to model/predict damping properties of the materials. Advanced FEM-based modelling and modal strain energy approaches [20] could be one way to achieve this. Systems that include statistical variation would reduce the sensitivity "optimised" solutions to parameters. Thus main goal of our joint team is the development, and validation of the computational tools and algorithms required to model/predict the damping properties of materials reinforced with nanoparticles/fibres/tubes etc. over broader operating ranges in frequency, amplitude and temperature.

3. Low-cost damping systems: while many of the novel materials give excellent performance, industries such as domestic appliances have a need for damping that is cheaper than current bitumen-based coatings. Research into the damping properties of low-cost materials is required; air and sand have already been used and require further application to nanoscale materials to yield significant damping.

The critical issues to be considered include: (a) choice between nanotubes and related materials (SWNTs and MWNTs, BN or SiC nanotube) for vibration damping; (b) tailoring the nanotube/matrix interface with respect to the matrix; (c) orientation and architecture of the nanotubes in matrix, its dispersion and the control of the nanotube/matrix bonding. Multiscale vibration damping modelling and computational simulation of materials with nanoscale reinforcement is still remaining a challenge.

## 3. MODELLING APPROACHES

In this paper, our idea stems from those carbon-based materials (nanotubes) have some similarities between the molecular model of a nanotube and the structure of a frame building. In a carbon nanotube, carbon atoms are bonded together by covalent bonds. These bonds have their characteristic bond lengths and bond angles in a three-dimensional space. Thus, it is logical to simulate the deformation of a nanotube based on the method of classical structural mechanics [11-15, 38]. In following sections, we first establish the bases of this concept and then demonstrate the approach by a few computational examples.

Basic equations of motion can determine the orientation of a single particle that is suspended in a viscoelastic medium such as polymeric material. The particle is regarded as a rigid body. The actual position of each point of a body, in relation to a frame of reference, is specified with the help of its position vector. The interaction between nanoparticles and matrix consists of two parts: the interpolation of the nanoparticle grid velocity v(x, t) onto the structural nodal velocities vs (xs, t) and the distribution of the nanoparticle structural interaction force f FSIs (Xs, t) onto the matrix f FSI (x,t) as shown in fig. 2.

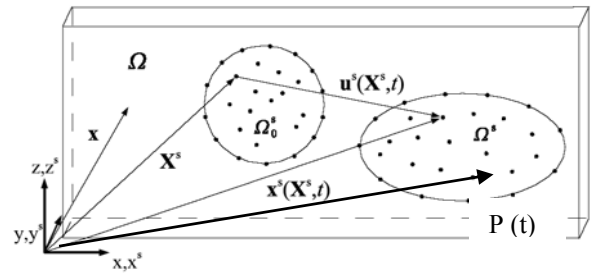

Fig. 2. Reference and actual position coordinates of the points of a rigid body

For the motion of a rigid body six degrees of freedom must be considered (e.g. [2]). In the case of cylindrical body with mass m, radius R and length H, tensor $C_0$ is transversally isotropic and can be shown in the form as follows [Holm]:

$$C_0 = \frac{m}{2} R^2 m_0 \otimes m_0 + \frac{m}{12}(3R^2 + H^2)(E - m_0 \otimes m_0)$$

The computational domain $\Omega$ are described with the time invariant position vector x; the solid positions in the initial configuration $\Omega s_0$ and the current configuration $\Omega s$ are represented by Xs and xs (Xs, t), respectively. The movement parameters of the particle are defined in the table 2.

Table 2: Nomenclature of the computational technique

|  | Carbon-based material Domain ($\Omega f$) | Solid Domain ($\Omega s$) |
|---|---|---|
| Spatial coordinate | x | $x^s$ |
| Displacement | - | $u^s = x^s - X^s$ |
| Velocity | v | $v^s = du^s/dt = \dot{u}^s$ |
| Acceleration | - | $a^s = d^2u^s/dt^2 = \ddot{u}^s$ |

It can be shown that the initial conditions of the components of rotation vector θ can be set as a function of time and thus the rotation tensor P(t) can be written as follows:

$$P(\theta) = \frac{1-\cos\theta}{\theta^2}\theta \otimes \theta + \frac{\sin\theta}{\theta}\theta \times E + \cos\theta E$$

and with the vector of angular velocity w(t)=P·Ω, the above equation gives the solution of equation of particle motion. m0 is initial position that could be specified and was chosen as follows:

$$m_0 = \frac{1}{\sqrt{3}}\begin{pmatrix}1\\1\\1\end{pmatrix}$$

In both cases, the classical solution and the above presented model of motion, the course of a rotating particle is similar to those shown in Figure 3 (marked with circles). If interactions between the particle and the surrounding medium occur, the tensor is C0 ≠ 0. It can be shown that, if interactions are taken into consideration, the motion of the particle is slowing down. In Figure 4 the line with circular symbols shows the motion if friction is neglected. The consideration of friction results in the line with triangular symbols.

The dynamic characteristics of the CNT-reinforced structure can be described by its eigenvalues and eigenvectors [23]. These establish the relation between loss factor, damping ratio and the structural deformation. The equilibrium of the damping structure can be described by the following equation in matrix form and then solved by FEM-based approach [9-13] for the 2 degree of freedom system:

$$M \cdot \Delta\ddot{D}^{n+1} + C^n \cdot \Delta\dot{D}^{n+1} + K^n \Delta D^{n+1} = \Delta F^{n+1} \quad (1)$$

where the superscript n and n+1 are the incremental steps; M is the mass matrix; Cn and Kn are the damping and stiffness matrices at the nth step, respectively. The vectors $\Delta D^{n+1}$ and $\Delta F^{n+1}$ are displacement and force at the (n+1)th step, respectively.

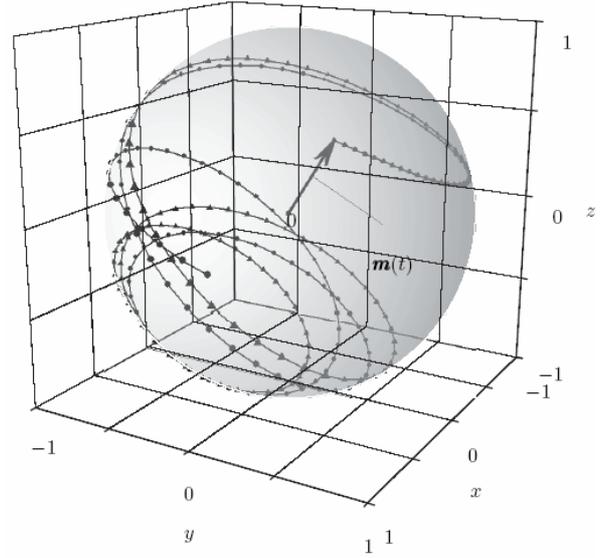

Fig. 3. Rotation track of the particle in the interval $0s \leq t \leq 5s$; circles without friction; triangles with friction

The average value of surface strain is used to represent structural deformation. Dissipated energy due to friction is equivalent to the frictional shear force the differential displacement between carbon fibre and a matrix [24]. The full energy dissipation, energy dissipation and loss factor for one loading cycle can be found respectively as follows:

$$U_{diss} = \int_V (\sigma_{ij}\varepsilon_{ij}/2)dV; \quad (2)$$

$$\Delta U = 2\tau^{max}(2\pi r_n l_n^2) \cdot (\varepsilon_z^m - \varepsilon_z^{max});$$

where indexes are related to m is matrix material, max is maximum displacement, n is nanoparticle, diss is dissipation energy, ε is strain, and τ is shear stress.

Substituting the matrices in the Eq. (1) by corresponding matrices given by the Eqs. (3), the fully integrated set of equations can be obtained for prediction of damping behaviour of the CNT-reinforced system. The stiffness matrix for the Eq. (1) has been established as shown in [9-13, 26]. The mass and damping matrices of the CNT-reinforced material (Fig. 1e) can be found with the following consideration. If the mass per unit length (m) is constant along the length of material, it can be taken out of the integral equation. The mass matrix of the material is fully populated. If loaded along the axes, Z and assumed two degree of freedom $\int\psi_b\psi_b dx=L$ [26], the matrices can be written in the symmetric form as follows:

$$M_2 = m_2 \begin{bmatrix} \int_{a_1}^{a_2} \psi_1 \psi_1 dx & \int_{a_1}^{a_2} \frac{1}{2} \psi_1 \psi_2 dx & \int_{a_1}^{a_2} \frac{(t_2 - 2t_1)}{4} \psi_1 \varphi^/ dx \\ - & \int_{a_1}^{a_2} \psi_2 \psi_2 dx & \int_{a_1}^{a_2} \frac{(2t_2 - t_1)}{3} \psi_b \psi_b dx \\ - & - & \int_0^L (\varphi\varphi + \frac{(t_1^2 - t_1 t_2 + t_2^2)}{4}\varphi^/ \varphi^/)dx \end{bmatrix};$$

$$C = \begin{bmatrix} a_1 E_x (1+\eta_{xi}) & a_1 E_x (1+\eta_{xi}) & a_1 E_x (1+\eta_{xi}) \\ a_1 E_y (1+\eta_{yi}) & a_1 E_y (1+\eta_{yi}) & a_1 E_y (1+\eta_{yi}) \\ a_1 E_z (1+\eta_{zi}) & a_1 E_z (1+\eta_{zi}) & a_1 E_z (1+\eta_{zi}) \\ 0 & 0 & 0 \\ 0 & 0 & 0 \\ 0 & 0 & 0 \end{bmatrix} \quad (3)$$

where ν is Poisson ratio in three-dimensional space (x, y, z), a is coefficient defined as follows:

$$a_1 = \frac{1 - v_1 v_2}{1 - v_3^2 - 2v_1 v_2 - 2v_1 v_2 v_3}; \quad a_2 = \frac{v_3 + v_1 v_2}{1 - v_3^2 - 2v_1 v_2 - 2v_1 v_2 v_3};$$

$$a_3 = \frac{1 - v_3}{1 - v_3 - 2v_1 v_2}; \quad a_4 = \frac{1 - v_1}{1 - v_3^2 - 2v_1 v_2}; \quad a_5 = \frac{1 - v_2}{1 - v_3^2 - 2v_1 v_2}$$

Damping behaviour of CNT-reinforced polymeric material can be investigated further at resonance frequencies between 10Hz and 10,000Hz band. The damping performance could be determined by Modal Strain Energy analysis [25] and laser vibrometry, and then compared to the predicted numbers by the Egs. (1-2). However, it will not be considered in this paper.

## 6. CONCLUSION

Nanoparticles/tubes can be used as a reinforcement of a matrix to provide multi-functionality, and thus we need to create an environment (knowledge) to introduce nanomaterials widely in industry. Commercial utilisation of a damping technology depends from both technical performance and business environment for that. This has often been seen to be a limiting factor in the utilisation of novel material/technology. The principal conclusions are that by invoking the properties of nano-auxetics/nanostructures it is possible to control the wave/sound/vibration propagation in the material and enhance the energy dissipation that can assist in improving the inherent damping of materials, but an extensive experimental/theoretical environment is required to apply it.

The mechanisms involved in such materials need to be understood and the relevance to damping identified.

The ongoing technology/material design multi-University project at the core of the concept is a nanoparticles/fibres/tubes-reinforced composite material and coating technology with regard to extensive dynamic characterization and modelling that includes two key elements: computational simulation and engineering workbench tools. The manufacturing design comprises several steps for tailoring the nanoparticle-matrix interface, dispersion and orientation including control over related technological parameters. Plausible aerospace/automotive-oriented damping materials of the next generation may be CNT-reinforced metallic foams, CNT-reinforced polymeric materials and CNT-reinforced ceramic coating materials. Selection of nanoparticle reinforcement depends on their price and production volume.

## 7. ACKNOWLEDGEMENTS

Support of the research work by the Royal Society in the United Kingdom and the WELCH scholarship administered through the Amer. Vac. Soc. / Int.-l Union for Vac. Sc., Tech. and Applications in the U.S.A. and Europe is gratefully acknowledged. Dr. Bochkareva is currently continuing her research work under the EU INTAS 2005-2007 postdoctoral fellowship Ref. Nr 04-83-3067. Further research work is also being supported by Marie-Curie Fellowship Ref. # 021298-Multiscale Damping at the Rolls-Royce Center in Damping Technology, the University of Sheffield in the U.K. It should be noted that the views expressed in this paper are those of the authors and not necessarily those of any institutions.